\documentstyle[amssymb,aps]{revtex}

\begin{document}
\title{Quantum-mechanical calculation of Stark widths of Ne\thinspace {\sc VII} $%
n=3,$ $\Delta n=0$ transitions}
\author{Yuri V. Ralchenko\thanks{%
Electronic mail: fnralch@plasma-gate.weizmann.ac.il}}
\address{Department of Particle Physics, Weizmann Institute of Science, Rehovot\\
76100, Israel}
\author{Hans R. Griem\thanks{%
Electronic mail: griem@glue.umd.edu}}
\address{Institute for Plasma Research, University of Maryland, College Park,\\
Maryland 20742}
\author{Igor Bray\thanks{%
Electronic mail: igor@yin.ph.flinders.edu.au} and Dmitry V. Fursa\thanks{%
Electronic mail: dmitry@yin.ph.flinders.edu.au}}
\address{Electronic Structure of Materials Centre, School of Physical Sciences, \\
The Flinders University of South Australia, G.P.O. Box 2100, Adelaide 5001,\\
Australia}
\maketitle

\begin{abstract}
The Stark widths of the Ne{\sc \thinspace VII} $2s3s-2s3p$ singlet and
triplet lines are calculated in the impact approximation using
quantum-mechanical Convergent Close-Coupling and Coulomb-Born-Exchange
approximations. It is shown that the contribution from inelastic collisions
to the line widths exceeds the elastic width contribution by about an order
of magnitude. Comparison with the line widths measured in a hot dense plasma
of a gas-liner pinch indicates a significant difference which may be
naturally explained by non-thermal  Doppler effects from persistent implosion
velocities or turbulence developed during the pinch implosion.
Contributions to the line width from different partial waves and types of
interactions are discussed as well.
\end{abstract}


\section{Introduction}

Spectral line shapes can provide very rich and valuable information on
important plasma parameters, such as ion and atom temperature, electron
density, electric field distributions, etc. The quantum-mechanical theory of
collisional impact line broadening is well established and developed \cite
{Griem74}, however, the number of purely quantum calculations, especially
for highly charged ions, is rather limited. Most theoretical efforts were
directed toward elaboration of semiclassical or semiempirical methods which
showed good accuracy for neutrals and low-charge ions. It is only recently,
when a number of sophisticated atomic collisional codes have become
available, that high-quality quantum-mechanical results could be applied to
line shape calculations for highly charged ions. From the experimental point
of view, test measurements of line profiles are impeded by the required
independent determination of plasma temperature and density. The
experimental situation is even more peculiar in that the line widths of
high-Z ions were measured almost exclusively by the Bochum group (see \cite
{Glenzer1996,Wrubel1998} and references therein), and therefore lack an
independent confirmation.

The recent results on the Stark broadening of spectral lines from multiply
charged ions revealed a significant discrepancy between the independent
quantum-mechanical calculations and, on the other hand, experimental
measurements and semiclassical results. For the B{\sc \thinspace III}
measurements \cite{GleKun96,Alexiou97}, the Stark line widths for the
simplest $2s-2p$ transition differ by as much as a factor of 2, the two
quantum results \cite{Griem1997,Seaton88} being in agreement to within 10\%.
A possible explanation for this discrepancy in terms of a developed
turbulence and different treatments of small partial waves in electron-ion
scattering was proposed in Ref. \cite{Griem1997}; however, more comparisons
and detailed investigation of important contributions to the line width are
of primary importance.

Measurements of line profiles for the $2s3s-2s3p$ transitions $%
^{1}S_{0}-^{1}P_{1}$ and $^{3}S_{1}-$ $^{3}P_{2}$ of Ne{\sc \thinspace VII}
emitted from a hot dense plasma of a gas-liner pinch were reported recently 
\cite{Wrubel1998,Wrubel1996b}. The experimental line widths for singlet $%
(\lambda ^{S}=3643.6$ \AA $)$\ and (the strongest) 
triplet $(\lambda ^{T}=1982.0$ \AA $)$\ 
lines are $\Delta \lambda ^{S}=1.70\pm 0.26$ \AA\ and $\Delta \lambda
^{T}=0.45\pm 0.07$ \AA ,\ respectively. The electron density and temperature
were measured {\em independently }by laser Thomson scattering and turned out
to be in the ranges $N_{e}=(3-3.5)\cdot 10^{18}$ cm$^{-3}$ and $%
T_{e}=(19-20.5)$ eV. The measured line widths agree well with most
semiclassical \cite{Griem74,Alexiou97,Alexiou1995} or semiempirical \cite
{HeyBre82} calculations but only marginally with other semiempirical results 
\cite{DK1981}.

Here we present the results of fully quantum-mechanical calculations of the
Stark line widths for the $n=3\rightarrow 3$ transitions in Be-like Neon.
The plan of this paper is as follows. In Sec. II the calculational method is
described. The features of atomic structure as well as inelastic and elastic
contributions to the line widths are discussed in detail. In Sec. III a
comparison with available experimental and theoretical results is made and
the sources of discrepancy are investigated. Finally, Sec. IV contains
conclusions and recommendations.

\section{Method}

\subsection{General theory}

The calculational method applied here is basically the same as the one used
in Ref. \cite{Griem1997}. We start from the fundamental formula for the full
collisional width at half-maximum (FWHM) for an isolated line corresponding
to a transition $u\rightarrow l$ \cite{Bar583}:

\begin{equation}
w=N_{e}\int\limits_{0}^{\infty }vF\left( v\right) \left( \sum_{u^{\prime
}\neq u}\sigma _{uu^{^{\prime }}}\left( v\right) +\sum_{l^{\prime }\neq
l}\sigma _{ll^{^{\prime }}}\left( v\right) +\int \left| f_{u}\left( \theta
,v\right) -f_{l}\left( \theta ,v\right) \right| ^{2}d\Omega \right) dv,
\label{Width_Bar}
\end{equation}
with $N_{e}$ being the electron density, $v$ the velocity of the scattered
electron, and $F\left( v\right) $ the Maxwellian electron velocity
distribution. The electron impact cross sections $\sigma _{uu^{^{\prime }}}$ 
$\left( \sigma _{ll^{^{\prime }}}\right) $ represent contributions from
transitions connecting the upper (lower) level with other perturbing levels
indicated by primes. In Eq. (\ref{Width_Bar}), the $f_{u}\left( \theta
,v\right) $ and $f_{l}\left( \theta ,v\right) $ are elastic scattering
amplitudes for the target ion in the upper and lower states, respectively,
and the integral is performed over the scattering angle $\theta $, with $%
d\Omega $ being the element of solid angle. Equation (\ref{Width_Bar})
relates a line width in the impact approximation with atomic cross sections,
facilitating the use of well-developed techniques of atomic scattering
calculations for line broadening studies. It can also be rewritten in terms
of the elastic {\bf S}-matrix elements assuming LS coupling (see, e.g., \cite
{Peach1996}):

\begin{eqnarray}
w &=&%
\mathop{\rm Re}%
\left\{ 2\pi N_{e}\sum\limits_{L_{u}^{T}L_{l}^{T}S^{T}ll^{^{\prime
}}}(-1)^{l+l^{^{\prime }}}\left( 2L_{u}^{T}+1\right) \left(
2L_{l}^{T}+1\right) 
{\displaystyle {(2S^{T}+1) \over 2\left( 2S+1\right) }}%
\right.  \nonumber \\
&&\times \left\{ 
\begin{array}{lll}
L_{l}^{T} & L_{u}^{T} & 1 \\ 
L_{u} & L_{l} & l
\end{array}
\right\} \left\{ 
\begin{array}{lll}
L_{l}^{T} & L_{u}^{T} & 1 \\ 
L_{u} & L_{l} & l^{\prime }
\end{array}
\right\} \int\limits_{0}^{\infty }%
{\displaystyle {1 \over v}}%
F\left( v\right) dv  \nonumber \\
&&\times \left. \left[ \delta _{l^{^{\prime }}l}-S_{U}\left( L_{u}Sl^{\prime
}%
{\textstyle {1 \over 2}}%
L_{u}^{T}S^{T};L_{u}Sl%
{\textstyle {1 \over 2}}%
L_{u}^{T}S^{T}\right) S_{L}^{*}\left( L_{l}Sl^{\prime }%
{\textstyle {1 \over 2}}%
L_{l}^{T}S^{T};L_{l}Sl%
{\textstyle {1 \over 2}}%
L_{l}^{T}S^{T}\right) \right] \right\} .  \label{Eq_Pe}
\end{eqnarray}
Here $L$ and $S$ are the atomic orbital angular momentum and spin, $l$ and $%
l^{\prime }$ are the electron orbital angular momentum before and after
collision, superscript $T$ denotes the quantum numbers of the total
electron+ion system, and the {\small $\left\{ 
\begin{array}{lll}
a & b & c \\ 
d & e & f
\end{array}
\right\} $} are 6-j symbols. The advantage of Eq. (\ref{Width_Bar}) is that
it gives more clear insight as to the importance of inelastic and elastic
contributions to the line width; therefore, we will mainly be referring to
Eq. (\ref{Width_Bar}) in what follows.

In the present work, the inelastic cross sections appearing in Eq. (\ref
{Width_Bar}) were calculated with two independent methods, i.e., the
Convergent Close-Coupling (CCC) and Coulomb-Born-Exchange (CBE)
approximations. The basic idea of the CCC method \cite{Bray94} lies in the
close-coupling expansion with a large number of square-integrable states. A
set of coupled Lippmann-Schwinger equations for the transition matrix is
solved in momentum space, and the convergence of the results may be checked
easily by increasing the number of the basis functions. The details of the
CCC method can be found in a number of recent reviews \cite{BrSt95,FurBr97},
where a very good agreement with various experimental collisional data is
shown. For the calculations with the more traditional CBE approximation we
made use of the code {\sc ATOM} described in Ref. \cite{SheVa93}. In
addition to the Coulomb attraction between ion and electron and exchange, 
{\sc ATOM} accounts for normalization (unitarization) effects and uses
experimental level energies when calculating the atomic wave functions and
collisional cross sections. It is well known that the Coulomb-Born
approximation corresponds to perturbation theory with $1/Z$ as expansion
parameter, where $Z$ is the spectroscopic charge; therefore, one can expect
better accuracy for the CBE method applied to {\em highly charged} ions.
Although the CCC method generally provides a superior accuracy, the use of 
{\sc ATOM} greatly reduces the computational efforts. Comparison of CCC and
CBE cross sections for highly charged H- and Li-like ions $\left( Z\lesssim
12\right) $ demonstrated an excellent agreement between these two methods
and available experimental data \cite{Fisher97}.

\subsection{Atomic Structure}

Before proceeding to the details of collisional calculations, it is worth
mentioning some features of the Ne{\sc \thinspace VII} atomic structure.
First, the current version of the CCC code utilizes the Hartree-Fock (HF)
frozen-core approximation for atomic wave functions. To study the validity
of this approach, we have made a comparison of wave functions calculated
with the full HF and HF frozen-core methods using the Cowan code \cite
{Cowan1981}. The agreement between the two sets of wave functions proved to
be very good, thereby justifying the use of the frozen-core approximation.
These calculations were also used to determine the root mean squared radii
of the lower $2s3s$ and upper $2s3p$ states, which were found to be $%
1.81a_{0}$ and $1.79a_{0}$ respectively, where $a_{0}=0.529\cdot 10^{-8}$ cm
is the Bohr radius. (Recall that the CBE {\sc ATOM} code constructs atomic
wave functions by solving the Schr\"{o}dinger equation with {\em experimental%
} energies rather than solving {\em ab initio} HF equations.) Another
measure indicating the level of accuracy are the oscillator strengths, which
for the CCC calculations are found to agree within one percent with the
Opacity Project results \cite{Opac1990}. Since in some cases the {\sc ATOM%
} oscillator strengths $f_{ATOM}$ deviate from the high-accuracy results $%
f_{acc}$ by as much as 15\%, the CBE dipole-allowed excitation cross
sections were rescaled by the ratio $f_{acc}/f_{ATOM}$ to improve these
results. Finally, to check the applicability of LS coupling, we carried out
large scale atomic structure calculations for Ne{\sc \thinspace VII} with
Cowan's code taking into account both intermediate coupling and
configuration interaction. The results obtained show that the levels of
interest of Ne{\sc \thinspace VII} correspond to practically pure LS
coupling, although configuration interaction is important for the $2s3p$ $%
^{1}P$ and $2p3s$ $^{1}P$ states which mix to a level of 10\%. Nonetheless,
this mixture is unlikely to be important, since in the sum of inelastic
cross sections in Eq. (\ref{Width_Bar}) this effect is essentially smoothed
out.

\subsection{Inelastic collisions}

The inelastic cross sections appearing in Eq. (\ref{Width_Bar}) include all
possible electron-induced transitions originating from the lower or upper
states of a transition. It is normally safe to neglect the ionization and
recombination processes, taking into account only electron impact excitation
and deexcitation. For the line widths discussed here, even the $\Delta n\neq
0$ excitations may (but will not) be ignored, since their rates are at least
two orders of magnitude smaller than those for the $\Delta n=0$ transitions.
Table I presents the CBE rate coefficients for electron impact excitation
and de-excitation processes connecting the upper and lower levels of
transitions with other perturbing $2l3l^{\prime }$ levels. The calculation
was carried out for an electron temperature $T_{e}=20$ eV, which corresponds
to the experimental conditions of Ref. \cite{Wrubel1996b}, and only
one-electron transitions are considered here since two-electron transitions
were found to have much smaller cross sections. One can see that the largest
rate coefficients correspond to dipole-allowed transitions, while
dipole-forbidden and spin-forbidden channels contribute only a few percent
to the inelastic part of the line width. It should also be noted that since
the reaction thresholds are smaller than 6-7 eV, the rate coefficients are
rather insensitive to small (a few eV) variations in the electron
temperature around the experimental value of 20 eV.

Both calculational methods give close (within 10\%) results for the most
important dipole-allowed cross sections. (Note that the excitation of the
inner $2l$ electron is also significant for the line width, contributing as
much as 12\% and 8\% for singlet and triplet lines, respectively.) An
example of the agreement between the CCC and CBE results is demonstrated on
Fig. 1, where $2s3s$ $^{1}S$ - $2s3p$ $^{1}P$ and $2s3p$ $^{3}P$ - $2s3d$ $%
^{3}D$ excitation cross sections are shown. Unfortunately, there are no
other available theoretical nor experimental data for the $3-3$ transitions
in Ne{\sc \thinspace VII}, so in order to test the accuracy of our
calculations it seems to be reasonable to make a comparison with the
existing $2-3$ data for this ion. Probably, the most accurate theoretical
results were produced recently by Ramsbottom {\em et al.} \cite{Rams95}, who
calculated electron impact excitation rates for many $2-3$ transitions using
the multichannel R-matrix method. The comparison shows very good agreement
between our data and those of Ref. \cite{Rams95}. For instance, the CBE
excitation rate coefficients ($T_{e}$ = 10$^{6}$ K $\approx 86$ eV) 
for the outer electron
transition $2s2p$ $^{3}P$ - $2s3d$ $^{3}D$ and inner electron transition $%
2s2p$ $^{3}P$ - $2p3s$ $^{3}P$ are $6.8\times 10^{-10}$ cm$^{3}$s$^{-1}$ and 
$9.1\times 10^{-11}$ cm$^{3}$s$^{-1}$, respectively, which agree well with
the R-matrix values of $6.2\times 10^{-10}$ and $9.6\times 10^{-11}$ cm$^{3}$%
s$^{-1}$. There also exist $\theta $-pinch experimental results \cite
{JohnKun71} for excitation rates from the ground and metastable states to
some of the $n=3$ states at an electron temperature of 260 eV; these are 2
to 3 times smaller than CCC/CBE rates, but large experimental errors up to
200-300\% limit their usefulness.

To summarize, for the experimental conditions of Ref. \cite{Wrubel1996b},
the CBE inelastic contribution (with {\em account} of the $\Delta n=1$
transitions) to the line widths obtained from Eq. (\ref{Width_Bar}) is $%
w_{in}^{S}\approx 0.806$ \AA\ for the singlet line and $w_{in}^{T}\approx
0.197$ \AA\ for the triplet line.

\subsection{Elastic Collisions}

According to Eq. (\ref{Width_Bar}), the non-Coulomb elastic amplitudes of
scattering from the upper and lower states at the same electron impact
energy should be subtracted and averaged over the Maxwellian electron energy
distribution. These amplitudes were calculated for a large range of electron
energies only with the CCC code, since the existing version of the CBE code 
{\sc ATOM} produces only inelastic cross sections. The $2s3s$ $^{3}S$ and $%
2s3p$ $^{3}P$ elastic cross sections $\sigma _{el}(E)$ along with the
coherent difference term $\tilde{\sigma}(E)$ $\equiv \int \left| f_{s}\left(
\theta ,v\right) -f_{p}\left( \theta ,v\right) \right| ^{2}d\Omega $ are
shown on Fig. 2, the singlet cross sections and difference term having a
similar behavior. These results unambiguously reveal the same peculiarities
as were noticed for the B{\sc \thinspace III} $2s-2p$ elastic term \cite
{Griem1997}, i.e., a faster than $1/E$ energy dependence and strong
cancellation in $\tilde{\sigma}(E)$. For example, at electron impact
energies $E\gtrsim 30$ eV, the coherent difference $\tilde{\sigma}$ is more
than an order of magnitude smaller than any of the $\sigma _{el}$. Since at
large energies the elastic cross section is mainly determined by the size of
a system, such a cancellation may be due to almost equal mean squared radii
of the $2s3s$ and $2s3p$ states, as was already mentioned above. The general
behavior of the elastic difference term deserves a special investigation and
will be reported elsewhere. The Maxwell-averaged elastic contribution to the
line width is $w_{el}^{S}\approx 0.067$ \AA\ and $w_{el}^{T}\approx $ $0.023$
\AA\ for singlet and triplet, respectively. This shows that in this case the
elastic contribution to the line widths is about an order of magnitude
smaller than the inelastic one, which is not surprising for such high
temperatures.

\subsection{Final results}

To summarize, the total line widths (FWHM) for the $2s3s-2s3p$, 
$^{1}S-^{1}P$
and $^{3}S-^{3}P$, transitions obtained from Eq. (\ref{Width_Bar}) are $%
w_{1}^{S}\approx 0.873$ \AA\ and $w_{1}^{T}\approx 0.220$ \AA . The same
widths were also calculated with Eq. (\ref{Eq_Pe}) using the CCC elastic 
{\bf T}-matrix elements and the relation between {\bf T}-matrix and {\bf S}%
-matrix ${\bf \hat{T}}={\bf \hat{S}}-{\bf \hat{I}}$ (${\bf \hat{I}}$ is the
unit matrix). The corresponding singlet and triplet widths are $%
w_{2}^{S}\approx 1.05$ \AA\ and $w_{2}^{T}\approx 0.230$ \AA . The
difference between the results obtained with Eqs. (\ref{Width_Bar}) and (\ref
{Eq_Pe}) can probably be attributed to the resonances in the CCC {\bf T}%
-matrix, which were not included into the CBE inelastic calculations. A
conservative estimate of the accuracy of these results, based on the CCC-CBE
agreement and the accuracy of the CCC calculations along the [Be] sequence,
is 15 \%. Thus, the final Stark line widths are

\begin{equation}
w^{S}=(1.0\pm 0.15)\text{ \AA },\ w^{T}=(0.23\pm 0.03)\text{ \AA }.
\label{results}
\end{equation}

\section{Discussion}

The line widths calculated here differ noticeably from the measured values
of Ref. \cite{Wrubel1996b} and most of the theoretical data. The ratios of
experimental to different theoretical Stark widths for the Ne{\sc \thinspace
VII} lines are presented in Table II. The methods cited there cover various
modifications of the semiclassical \cite{Griem74,Alexiou1995} and
semiempirical \cite{HeyBre82,DK1981} approximations. The semiclassical
methods, including the latest nonperturbative calculations \cite{Alexiou1995}%
, yield values which are generally in agreement with the experimental data.
The semiempirical results of Dimitrijevi\'{c} and Konjevi\'{c} \cite{DK1981}
are rather close to our values, and this is quite similar to what had
already been noticed for the B{\sc \thinspace III} calculations 
\cite{Griem1997}.

The major diagnostics challenge in the gas-liner pinch experiment \cite
{Wrubel1996b} may be the determination of the main plasma parameters, i.e.,
the electron temperature and density, in a region where the multiply charged
ions of neon are situated. In the experiment both $T_{e}$ and $N_{e}$ were
determined from the Thomson scattering only {\em globally} which may not be
characteristic of the plasma conditions near the locally injected neon. As a
matter of fact, there exist some experimental indications that density and
temperature do vary in the vicinity of the doping gas \cite{Kunze98priv}.
However, the experimental value of $T_{e}$ is supported by the fact that
electron temperatures $T_{e}=19-20$ eV are well within the range of the
maximal abundance temperatures for Ne{\sc \thinspace VII} at an electron
density $N_{e}=\left( 3\div 4\right) \cdot 10^{18}$ cm$^{-3}$. Our
calculations with the collisional-radiative code {\sc NOMAD} \cite{Ral98}
show that for equilibrium conditions the Ne{\sc \thinspace VII} ions account
for about 30\% of the total amount of neon. Another line broadening
mechanism affecting the observed widths may be unresolved Doppler line
splitting associated with the radial implosion velocities in the gas-liner
pinch \cite{Kunze98priv}. The contribution from an ion (proton) collisional
broadening may be estimated using Eq. (517) from Ref. \cite{Griem74}, and it
is easy to show that ion broadening is negligibly small comparing to
electron impact broadening.

Since the experimental conditions in the Ne{\sc \thinspace VII} measurements
were basically the same as for the B{\sc \thinspace III} experiment, the
general conclusions \cite{Griem1997} regarding a possible effect of a
developed turbulence on the line widths should remain essentially the same.
It was mentioned in Ref. \cite{Wrubel1996b} that the measured value of Stark
width for the triplet transition $\Delta \lambda \approx 0.45$ \AA\
constitutes about 70\% of the total measured line width\footnote{%
There is no information in Ref. \cite{Wrubel1996b} on the full line width of
the singlet transition, so we will not discuss it in what follows.} which
therefore is $\Delta \lambda _{\exp }\sim 0.64$ \AA . This full width
includes Stark, Doppler and instrumental broadening, the latter being
decomposed into Gaussian (0.07 \AA ) and Lorentzian (0.05 \AA ) parts \cite
{Glenzer1996}. For an ion temperature of $T_{i}=T_{e}=20$ eV, the pure
Doppler width is approximately $\Delta \lambda _{D}\approx 0.15$ \AA . As
noted in \cite{Griem1997}, the Reynolds numbers for the Bochum gas-liner
pinch experiment are of the order of $10^{4}$, which is sufficient for a
developed turbulence to exist. Such a turbulence leads to an extra chaotic
motion of Ne ions with a characteristic velocity of the order of the proton
thermal velocity $v_{p}$. Hence, the full thermal+turbulent Doppler width
becomes a factor $\sqrt{20+1}\approx 4.6$ larger (here $20$ is the ratio of
masses $M_{Ne}/M_{H})$ and is now $\Delta \lambda _{D}\approx 0.70$ \AA .
Using Eq. (6) of Ref. \cite{Whiting68}, for the FWHM of a combined Voigt
profile including Stark, thermal+turbulent Doppler and instrumental
contributions, we get a value $\Delta \lambda \approx 0.85$ \AA\ which is
30\% higher than $\Delta \lambda _{\exp }$. The main uncertainty in this
calculation obviously comes from the turbulent contribution, which is rather
sensitive to the value of the characteristic velocity. It is straightforward
to show that reducing this velocity by one third only, i.e., multiplying the
pure Doppler width by 3.1 instead of 4.6, one can exactly reproduce the
experimental value of $\Delta \lambda _{\exp }$. Thus, according to the
hypothesis proposed in Ref. \cite{Griem1997}, reasonable values of
characteristic turbulent velocities may naturally explain the observed
difference in line widths.

Regarding the discrepancy between the quantum and other theoretical
calculations, the reader may wonder as to the source of such a difference.
The crucial point is that unlike the quantum-mechanical methods, the
semiclassical approaches have a natural limit of applicability arising from
the Heisenberg uncertainty principle (see, e.g., \cite{Williams45}). The
criterion of applicability of the semiclassical calculations may be
formulated \cite{Griem98} as a requirement for the distance of the closest
approach $r_{\min }$, rather than the impact parameter $\rho $, to be larger
or at least of the same order than the corresponding de Broglie wavelength, $%
\lambda _{\min }=2\pi \hbar /mv_{\max }$. Using the angular momentum
conservation it is straightforward to show that this is equivalent to the
inequality:

\begin{equation}
2\pi \lesssim L.  \label{2piL}
\end{equation}
Another limitation on impact parameters was introduced in order to avoid
violations of unitarity \cite{Griem74}, but still assuming the long-range 
dipole interaction to remain valid. Again reformulated in terms of the
distance of the closest approach the corresponding condition may be written 
as

\begin{equation}
\frac{r_{n}}{r_{\min }}\lesssim 1,  \label{unit}
\end{equation}
where $r_{n}$ is the excited state atomic radius. If this inequality is
violated, both semiclassical and long-range interaction approximation are
questionable. Using the Coulomb parameter $\eta =(Z-1)e^{2}/\hbar v,$ Eq. (%
\ref{unit}) may also be expressed in terms of the total angular momentum $L$
as \cite{Griem98}:

\begin{equation}
\frac{r_{n}}{r_{\min }}=\frac{(Z-1)\ r_{n}\left( \left[ 1+\left( L/\eta
\right) ^{2}\right] ^{1/2}+1\right) }{a_{0}L^{2}}\lesssim 1  \label{rn_rmin}
\end{equation}
As was noted above, the mean root squared radii of the $2s3s$ and $2s3p$
states are about 1.8\/\/$a_{0}$. For $T_{e}=20$ eV, the Coulomb parameter is 
$\eta $ $\simeq 7$ and therefore the ratio $r_{n}/r_{\min }$ takes values of
1.45, 0.96, 0.43 and 0.22 for $L=$ 4, 5, 8 and 12 respectively. It follows
then that for the given electron temperature, criteria (\ref{2piL}) and (\ref
{unit}) are similarly restrictive for the semiclassical approximation.

Unlike to the semiclassical method, in fully quantum-mechanical calculations
the determination of the range of significant $L$-values is naturally
accomplished by the partial wave expansion. In Fig. 3 the contribution of
different total electron+ion angular momenta $L_{T}$ to the CCC cross
sections is shown for an incident electron impact energy of 20 eV for a
number of transitions\footnote{%
The CBE partial wave composition is practically the same.}. Naturally, the
elastic cross sections are governed by the smallest values of $L_{T}$, which
are concealed in the strong collision term of semiclassical calculations.
The most important inelastic cross sections having the smallest thresholds
reach 50\% of their values only for $L=9$ for which the l.h.s. ratio of Eq. (%
\ref{unit}) is about $0.35$. This number is probably already sufficiently
small to justify the use of the long-range interaction approximation for $%
L\geq 9$; however, the restrictions following from Eq. (\ref{2piL}) are less
obvious to have been overcome.

Another discrepancy may come from other than dipole interactions. Although
the monopole interaction was not explicitly included into nonperturbative
semiclassical calculations \cite{Alexiou1995}, the quadrupole transition $%
2s3s-2s3d$ was shown to account for about 15\% of the line width. This value
is in contradiction to the present results. As one can see from Table I, the 
$2s3s-2s3d$ quadrupole channel contributes only approximately 3\% to the
quantum-mechanical inelastic line width. If we take into account only those
transitions that were considered in Ref. \cite{Alexiou1995}, then this
number increases to 3.5\%, still a factor of 4 smaller than the
nonperturbative semiclassical result.

These considerations clearly show that the accuracy of the semiclassical
calculations may not be as high as it is often thought to be, and new
calculations, both semiclassical and quantum-mechanical, are needed to
better establish the limits of applicability for the non-quantum methods.

\section{Conclusion}

A fully quantum-mechanical calculation of the Stark line widths for the
singlet and triplet $2s3s-2s3p$ lines of Ne{\sc \thinspace VII} was carried
out in the impact approximation with the use of accurate atomic data.
Although the results obtained disagree with experimental and most
theoretical results, a natural explanation for this disagreement can be
suggested. On one hand, the measurements are not free from difficulties
related to possible extra contributions from turbulence and unresolved
Doppler shifts. This suggests an independent measurement of Stark widths of
highly charged ions. On the other hand, the semiclassical calculations, not
obviously producing accurate results for other than dipole interactions, may
have problems when being applied to the small impact parameter region. In
our opinion, the next important step in the development of Stark broadening
theory would be a very detailed comparison between quantum and semiclassical
results.

\section{Acknowledgments}

This work was supported in part by the Israeli Academy of Sciences and the
Ministry of Sciences of Israel (Yu.V.R.), by the US National Science Foundation
(H.R.G.) and by the Australian Research Council (I.B. and D.V.F.).

\newpage

\begin{center}
{\bf Figure Captions}
\end{center}

{\bf Fig. 1.} Electron impact excitation cross sections for the transitions $%
2s3s$ $^{1}S$ - $2s3p$ $^{1}P$ and $2s3p$ $^{3}P$ - $2s3d$ $^{3}D$ in Ne{\sc %
\thinspace VII}. CBE - dashed lines, CCC - solid circles.

{\bf Fig. 2.} Non-Coulomb elastic cross sections of Ne {\sc VII} ions in $%
2s3s$ $^{3}S$ (solid line) and $2s3p$ $^{3}P$ (dot-dashed line) states, and
the coherent difference term $\tilde{\sigma}(E)$ (diamonds).

{\bf Fig. 3.} Contribution of different total electron+ion angular momenta $%
L_{T}$ to various elastic and inelastic cross sections.

\medskip 
\begin{table}[tbp] \centering%
\caption{The CBE electron impact excitation and de-excitation rate coefficients for Ne {\sc VII} 
in units of cm$^{3}$s$^{-1}$ for T$_{e}$ = 20 eV.\label{TableI}} 
\begin{tabular}{c|cccc}
\hline
& $2s3s\text{ }^{3}S$ & $2s3s\text{ }^{1}S$ & $2s3p\text{ }^{1}P$ & $2s3p%
\text{ }^{3}P$ \\ \hline
$2s3s\text{ }^{3}S$ & - & 7.46(-10) & 3.29(-10) & 6.10(-08) \\ 
$2s3s\text{ }^{1}S$ & 2.21(-10) & - & 5.17(-08) & 1.13(-10) \\ 
$2s3p\text{ }^{1}P$ & 2.47(-10) & 1.31(-07) & - & 5.30(-10) \\ 
$2s3p\text{ }^{3}P$ & 1.35(-07) & 8.44(-10) & 1.56(-09) & - \\ 
$2s3d\text{ }^{3}D$ & 9.96(-09) & 1.50(-09) & 1.48(-09) & 7.75(-08) \\ 
$2s3d\text{ }^{1}D$ & 4.10(-10) & 9.97(-09) & 6.45(-08) & 4.49(-10) \\ 
$2p3s\text{ }^{3}P$ & 9.71(-09) & 3.86(-10) & - & - \\ 
$2p3s\text{ }^{1}P$ & 9.99(-11) & 2.36(-08) & - & - \\ 
$2p3p\text{ }^{1}P$ & - & - & 1.20(-08) & 4.34(-11) \\ 
$2p3p\text{ }^{3}D$ & - & - & 2.04(-10) & 9.69(-09) \\ 
$2p3p\text{ }^{3}S$ & - & - & 3.89(-11) & 2.18(-09) \\ 
$2p3p\text{ }^{3}P$ & - & - & 1.12(-10) & 3.86(-09) \\ 
$2p3p\text{ }^{1}D$ & - & - & 7.04(-10) & 5.92(-11) \\ 
$2p3p\text{ }^{1}S$ & - & - & 6.63(-10) & 1.07(-11)
\end{tabular}
\end{table}%

\begin{center}
\begin{table}[tbp] \centering%
\caption{Ratio of the experimental Stark widths 
of the $2s3s-2s3p$ lines in Ne {\sc VII} to
different theoretical widths.\label{TableII}} 
\begin{tabular}{cc|c|c|c|c|c|c}
\hline
Line & ${\bf T}_{e}$ (eV) & ${\bf N}_{e}$ (cm$^{-3}$) & \multicolumn{5}{c}{$%
{\bf w}_{\exp }{\bf /w}_{theor}$} \\ \hline
Ne VII $^{1}S_{0}-^{1}P_{1}$ & 19 & 3.5$\cdot 10^{18}$ & 1.28$^{a}$ & 1.15$%
^{b}$ & 1.57$^{c}$ & 0.88(0.77)$^{d}$ & 1.70$^{e}$ \\ \hline
Ne VII $^{3}S_{1}-^{3}P_{2}$ & 20.5 & 3.0$\cdot 10^{18}$ & 1.53$^{a}$ & 1.29$%
^{b}$ & 1.91$^{c}$ & 0.94(0.82)$^{d}$ & 1.96$^{e}$ \\ \hline
\end{tabular}
\end{table}%
\end{center}

$^{a}$Semiclassical \cite{Griem74}, $^{b}$semiempirical \cite{HeyBre82}, $%
^{c}$semiempirical \cite{DK1981}, $^{d}$semiclassical \cite{Alexiou97}, $^{e}
$present work.


\begin{references}
\bibitem{Griem74}  H.R.Griem, {\em Spectral Line Broadening by Plasmas }%
(Academic, New York, 1974)

\bibitem{Glenzer1996}  S.Glenzer, in{\em \ Atomic Processes in Plasmas}, AIP
Conference Proceedings No. 381, edited by A.L.Osterheld and W.H.Goldstein
(AIP Press, New York, 1996), pp.109-122.

\bibitem{Wrubel1998}  Th.Wrubel, I. Ahmad, S.B\"{u}scher, H.-J.Kunze, and
S.H.Glenzer, Phys. Rev. E {\bf 57}, 5972 (1998).

\bibitem{GleKun96}  S.Glenzer and H.-J.Kunze, Phys.Rev. A {\bf 53}, 2225
(1996).

\bibitem{Alexiou97}  S.Alexiou, in {\em 13th International Conference on
Spectral Line Shapes}, AIP Conference Proceedings No. 386, edited by M.Zoppi
and L.Ulivi (AIP\ Press, New York, 1997), pp.79-98.

\bibitem{Griem1997}  H.R.Griem, Yu.V.Ralchenko, and I.Bray, Phys. Rev. E 
{\bf 56}, 7186 (1997).

\bibitem{Seaton88}  M.J.Seaton, J.Phys. B {\bf 21}, 3033 (1988).

\bibitem{Wrubel1996b}  Th.Wrubel, S.Glenzer, S.B\"{u}scher, H.-J.Kunze, and
S.Alexiou, Astron. Astrophys. {\bf 306}, 1023 (1996).

\bibitem{Alexiou1995}  S.Alexiou, Phys.Rev.Lett. {\bf 75}, 3406 (1995).

\bibitem{HeyBre82}  J.D.Hey and P.Breger, S. Afr. J. Phys. {\bf 5}, 111
(1982).

\bibitem{DK1981}  M.S.Dimitrijevi\'{c} and N.Konjevi\'{c}, in {\em 5th
International Conference on Spectral Line Shapes}, edited by B.Wende (Walter
de Gruyter, Berlin, 1981), pp.211-239.

\bibitem{Bar583}  M.Baranger, Phys.Rev. {\bf 112}, 855 (1958).

\bibitem{Peach1996}  G.Peach, in {\em Atomic, Molecular \& Optical Physics
Handbook}, edited by G.W.F.Drake (AIP Press, New York, 1996), Ch. 57.

\bibitem{Bray94}  I.Bray, Phys.Rev. A {\bf 49}, 1066 (1994).

\bibitem{BrSt95}  I. Bray and A.T.Stelbovics, Adv.At.Mol.Opt.Phys. {\bf 35},
209 (1995).

\bibitem{FurBr97}  D.V.Fursa and I.Bray, J. Phys. B {\bf 30}, 757 (1997).

\bibitem{SheVa93}  V.P.Shevelko and L.A.Vainshtein, {\em Atomic Physics for
Hot Plasmas} (IOP Publishing, Bristol, 1993).

\bibitem{Fisher97}  V.I.Fisher, Yu.V.Ralchenko, V.A.Bernshtam, A.Goldgirsh,
Y.Maron, H.Golten, L.A.Vainshtein, and I.Bray, Phys. Rev. A {\bf 55}, 329
(1997); V.I.Fisher, Yu.V.Ralchenko, V.A.Bernshtam, A.Goldgirsh, Y.Maron,
L.A.Vainshtein, and I.Bray, Phys. Rev. A {\bf 56}, 3726 (1997).

\bibitem{Cowan1981}  R.D.Cowan, {\em The Theory of Atomic Structure and
Spectra} (University of California Press, Berkeley, 1981).

\bibitem{Opac1990}  J.A.Tully, M.J.Seaton, and K.A.Berrington, J.Phys. B 
{\bf 23}, 3811 (1990).

\bibitem{Rams95}  C.A.Ramsbottom, K.A.Berrington, and K.L.Bell, At. Data
Nucl. Data Tables {\bf 61}, 105 (1995).

\bibitem{JohnKun71}  W.D.Johnston and H.-J.Kunze, Phys.Rev. A {\bf 4}, 962
(1971).

\bibitem{Kunze98priv}  H.-J. Kunze, private communication; see also
S.B\"{u}scher, Th.Wrubel, I.Ahmad and H.-J. Kunze, in {\em 14th
International Conference on Spectral Line Shapes}, AIP Conference
Proceedings (AIP\ Press, New York, 1998), to be published.

\bibitem{Ral98}  Yu.V.Ralchenko, V.I.Fisher and Y.Maron (unpublished).

\bibitem{Whiting68}  E.E.Whiting, J.Quant.Spectrosc.Radiat.Transfer {\bf 8},
1379 (1968).

\bibitem{Williams45}  E.J.Williams, Rev.Mod.Phys. {\bf 17}, 217 (1945).

\bibitem{Griem98}  H.R.Griem, in {\em 14th International Conference on
Spectral Line Shapes}, AIP Conference Proceedings (AIP\ Press, New York,
1998), to be published.
\end{references}
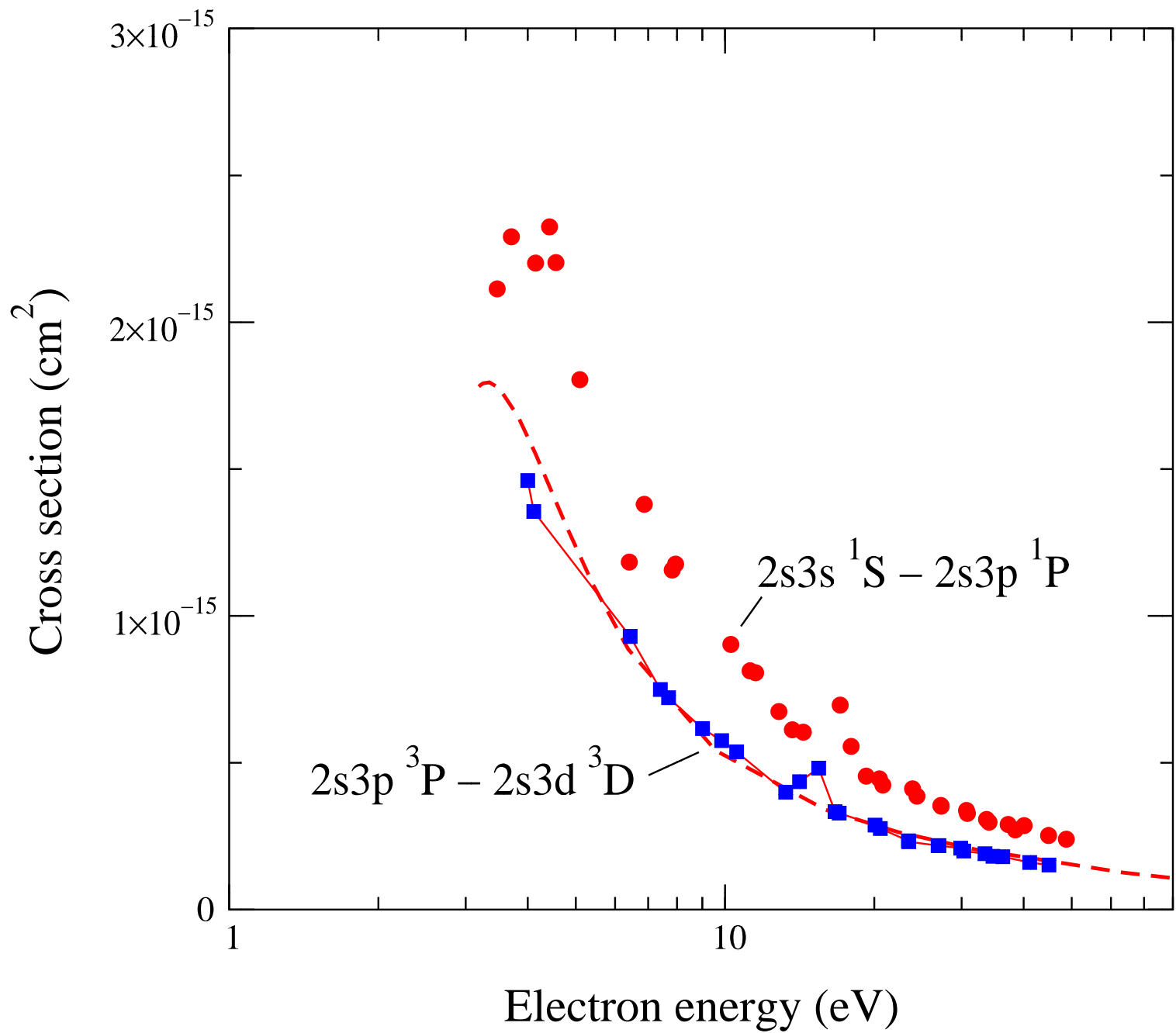
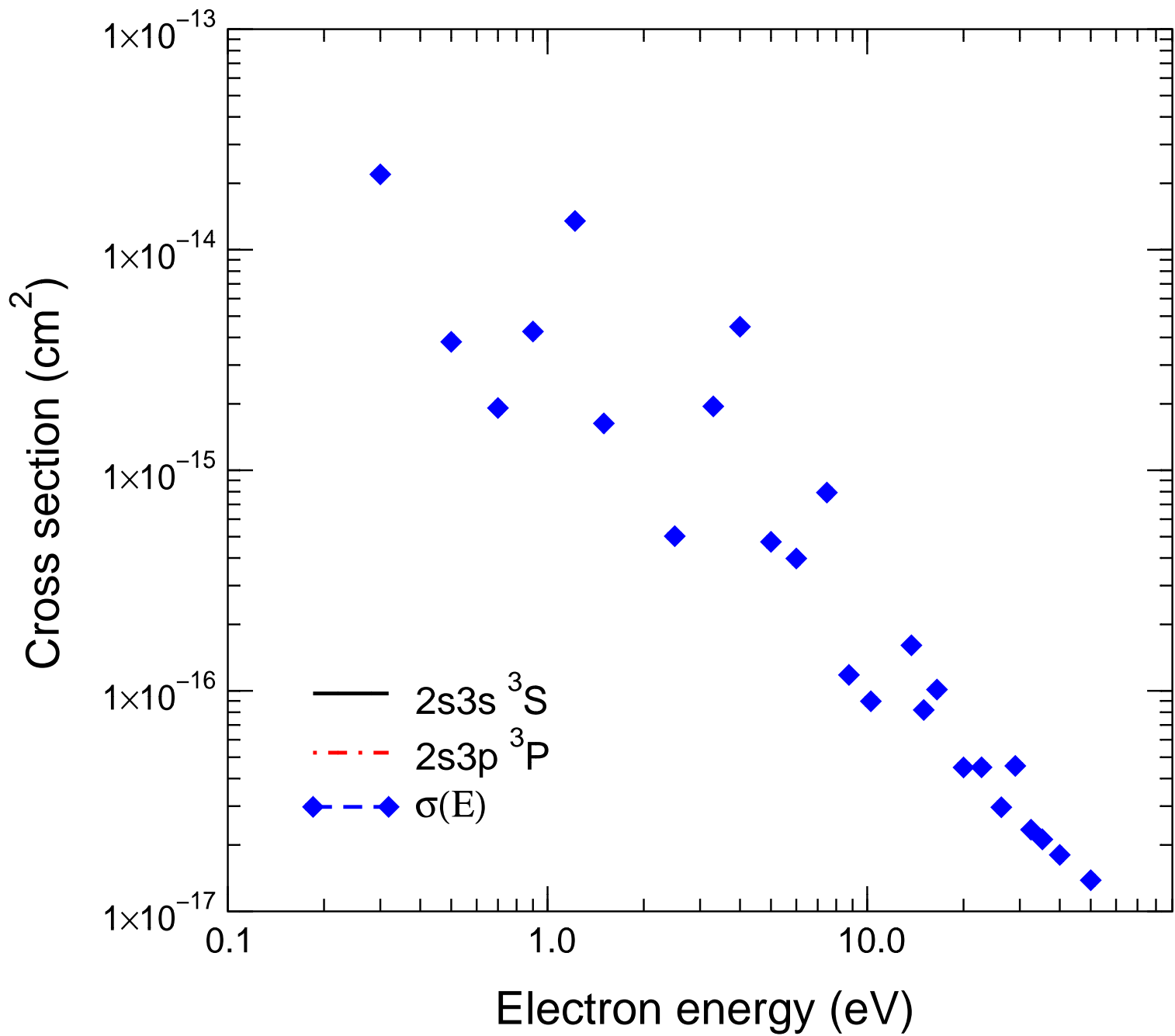
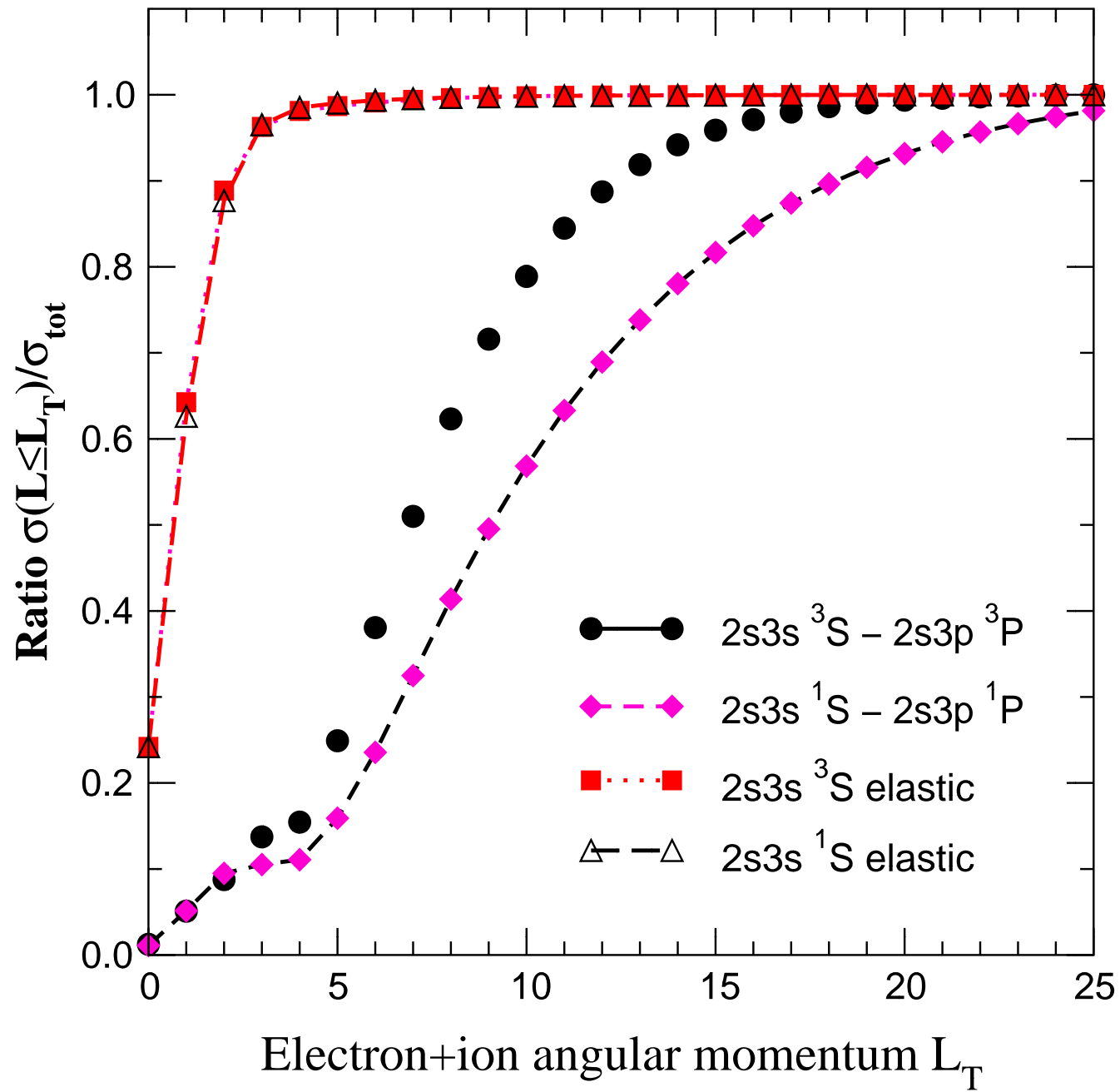
\end{document}